\documentclass[prb,twocolumn,showpacs,preprintnumbers,amsmath,amssymb,superscriptaddress]{revtex4-1}

\usepackage{graphicx}
\usepackage{dcolumn}
\usepackage{color}

\allowdisplaybreaks
\newcommand{\beq}{\begin{equation}}
\newcommand{\eeq}{\end{equation}}
\newcommand{\beqa}{\begin{eqnarray}}
\newcommand{\eeqa}{\end{eqnarray}}

\begin{document}
\title{Quantum dynamical screening of the local magnetic moment in Fe-based superconductors}

\author{A.~Toschi}
\affiliation{Institute for Solid State Physics, Vienna University of Technology 1040 Vienna, Austria}

\author{R.~Arita}
\affiliation{Department of Applied Physics, University of Tokyo, Tokyo 153-0064, Japan\\
JST, PRESTO, Kawauchi, Saitama 332-0012, Japan\\
JST, CREST, Hongo, Tokyo 113-8656, Japan}

\author{P.~Hansmann}
\affiliation{Institute for Solid State Physics, Vienna University of Technology 1040 Vienna, Austria}
\affiliation{Centre de Physique Th\'eorique, \'Ecole Polytechnique, CNRS-UMR7644, F-91128 Palaiseau, France}

\author{G.~Sangiovanni}
\affiliation{Institute for Solid State Physics, Vienna University of Technology 1040 Vienna, Austria}
\affiliation{Institut f\"ur Theoretische Physik und Astrophysik, Universit\"at W\"urzburg, Am Hubland, D-97074 W\"urzburg, Germany}

\author{K.~Held}
\affiliation{Institute for Solid State Physics, Vienna University of Technology 1040 Vienna, Austria}


\begin{abstract}
We have calculated the  local magnetic susceptibility of one of the prototypical Fe-based superconductors (LaFeAsO) by means of the local density approximation + dynamical mean field theory as a function of both (imaginary) time and real frequencies with and without vertex corrections. Vertex corrections are essential for obtaining the correct  $\omega$-dependence, in particular a pronounced low-energy peak at $\omega \sim 0.2\,$eV,  which constitutes the hallmark of the dynamical screening of a large instantaneous magnetic moment on the Fe atoms. In experiments, however, except for the case of x-ray absorption spectroscopy (XAS), the magnetic moment or the susceptibility represent typically the average over long time scales. In this respect, the frequency range of typical neutron experiments would be too limited to directly estimate the magnitude of the short-time moment.
\end{abstract}
\date{\today}
\maketitle

The discovery of high temperature superconductivity in iron based pnictides 
 \cite{kamihara08} not only gave new hope for increasing the critical temperatures towards room temperature but also posed new challenges for the understanding of unconventional superconductors, even regarding the properties of their normal and magnetic phases. It is generally accepted that conventional superconductors and also MgB$_2$ are weakly correlated electron systems, while cuprate superconductors with a magnetic insulating ground state, characterized by the opening of a Mott-Hubbard gap, are strongly correlated. In contrast, the situation in Fe-based superconductors is much less clear. The phase diagram is strikingly similar to that of cuprates with an antiferromagnetic parent compound and a  superconducting dome in the doped system. However, a major difference from the cuprate physics is certainly the metallicity of the antiferromagnetic spin-density-wave phase and of the low-doped normal region. Also the rather good performance of {\sl ab initio} (LSDA) calculations to describe\cite{mazin08,lilia2011} atomic positions and the symmetry of the long-range magnetic order of many Fe-based compounds may be interpreted as indicators for the irrelevance of electronic correlations in this compounds.

In fact, several groups have used weakly correlated theories to model iron based superconductors. These include the local (spin) density approximation (LDA, LSDA), and weak coupling perturbation theory such as the random phase approximation (RPA) and the fluctuation exchange (FLEX) approximation on top of LDA bandstructures \cite{kuroki09,ikeda10,arita09}. The latter show instabilities towards magnetism and superconductivity originating from nesting  vectors between different Fermi surface sheets. 

A first hint that a weakly correlated picture is, however, insufficient came from angular resolved photoemission spectroscopy (ARPES) experiments \cite{Malaeb08,Qazilbash09} showing that the LDA bandstructure needs to be renormalized by an effective mass enhancement of about 2 in LaFeAsO.  From the theoretical point of view this necessitates the weak coupling schemes to start from a correspondingly renormalized LDA bandstructure. Such a procedure is particular cumbersome for a multi-band system such as iron pnictides with five $d$ bands which are all renormalized somewhat differently and where electronic correlations also shift the orbitals relative to each other. In this situation a true many-body calculation is almost unavoidable, and indeed several groups have performed such calculations.
In particular, supplementing the LDA by dynamical mean field theory (LDA+DMFT) \cite{DMFTreview,LDADMFTrev} makes it possible to  describe well the one-particle spectrum and renormalization factors  \cite{haule08,craco08,anisimov09,aichhorn09,skornyakov09,ishida10,hansmann10,Yin11,Ferber11}.
A second point indicating that standard band-structure theories do not work is the ordered magnetic moment, which is for most compounds much larger in LSDA ($\approx 2 \mu_B$) than in experiment and hardly varies for different Fe-based superconductors, in contrast to experiment. In fact, the experimental values range from from 0.3-0.6$\mu_B$  in LaFeAsO (1111 compound) \cite{ishida09,Braden10} via 0.9$\mu_B$  in BaFe$_2$As$_2$  (122 compound) \cite{Huang08}  to 2.2$\mu_B$ in FeTe (11 compound) \cite{ishida09}. It is worth also noting here that  one generally expects the LSDA results to underestimate (rather than overestimate) the magnetic moments for strongly correlated electron systems.

Hence, there is evidently a big {\em moment puzzle} with theory predicting a large magnetic moments and experiments  measuring smaller ones for most of the Fe-pnictide compounds. In a recent Letter \cite{hansmann10}, we put forward a solution to this puzzle by considering local spin quantum fluctuations, which are active also above the magnetic ordering temperature. In fact, we showed that if one considers the local spin-spin correlation function of the paramagnetic phase of the Fe-based superconductors on very short time scales (fs) a relatively large local moment is observed, i.e., comparable to (or even larger than) the LSDA prediction of $2\mu_B$ for the ordered moment. On longer time scales, however, if the electron mobility is high enough, this local moment fluctuates very fast so that the time-averaged magnetic moment is considerably reduced. Such strongly screened moment is the one which can become magnetically ordered at low temperatures,  explaining the reduced size measured in neutron scattering experiments. Note that in addition to these temporal (but local in space) spin fluctuations, at low-temperature also non-local spatial (e.g. antiferromagnetic) spin fluctuations are expected to become relevant. In this respect, only the latter ones are included in the framework of an extended Heisenberg model, which has been recently analyzed in the context of neutron  experiments \cite{antropov} for iron-based superconductors. 

From a more theoretical perspective, the origin of the physics of large, but dynamically screened, local magnetic moments, as it emerges from our LDA+DMFT calculations,  can be related to the presence of several almost degenerate  moderately correlated Fe-bands at the Fermi level. This situation enhances the effects of the Hund's coupling interaction\cite{haule09}, which mainly controls the formation of the large local magnetic moment in a still rather itinerant metallic background, well fitting to the expression ''Hund's metals'' introduced in this context in Ref. \onlinecite{Yin11}.  

Based on our paramagnetic calculations and the itinerant nature of the antiferromagnetic phase, we suggested that the same mechanism not only reduces the local moment on long time scales in the paramagnetic phase but also the ordered moment in the antiferromagnetic phase, corresponding to an average over long time scales.
This was later confirmed by Yin, Haule and Kotliar \cite{Yin11,yin2} and, independently, by Misawa, Nakamura and Imada \cite{MNI2011}, who found a remarkably good agreement with the aforementioned experimental moments for different Fe-based superconductors (see also Ref. \onlinecite{Lee11}). 
The observed  material dependence of the ordered magnetic moments comes likely from the different degree of itinerancy of the several compounds, which results in a different degree of dynamical screening of the (almost material independent) short-time moment. Note that, in the LDA+DMFT framework, another explanation for the magnetic moment mismatch has been recently proposed by Hunpyo Lee et al. in  Ref. \onlinecite{Hunpyo2010}.

In this paper, we not only extend our earlier study \cite{hansmann10} for different temperatures and provide more details about our LDA+DMFT calculation for LaFeAsO, but we also present new results for the spectral properties of the (local) spin-correlations functions, aiming to analyze how  the hallmarks of the dynamically screened local moment can be traced in the existing experimental data. 

Specifically, Section \ref{Sec:method} describes how we calculate the temporal magnetic moment fluctuations via the spin-spin correlation function. In Section 
 \ref{Sec:Tdep}, we present new results for the temperature dependence of the spin-spin correlation function. In Section \ref{Sec:omega} we analytically continue the susceptibility  to the real frequency axis and discuss the importance of vertex corrections.
Section \ref{Sec:neutron} is devoted to a detailed comparison and test of the main result of our theoretical calculations with experiments. We analyze to what extent experimental findings of x-ray absorption  and neutron scattering spectroscopy are compatible with the existence of larger magnetic moments on short (fs) time scales, provided an high-frequency extrapolation of the experimental data is considered. Finally, we summarize the main results and our conclusions in Section \ref{Sec:Summary}.

\section{Method and Model}
\label{Sec:method}

\begin{figure}
\begin{center}
\includegraphics[width=0.75\columnwidth]{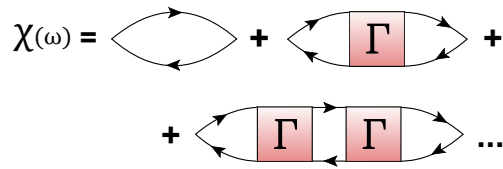}\\[0.1cm]
\end{center}
\caption{Diagrammatically, the local susceptibility $\chi$ computed in DMFT can be expressed in terms of the two (DMFT) Green functions, {\sl``bubble''} contribution (first diagram on the r.h.s.), plus vertex ($\Gamma$) corrections in a given particle-hole (in our case: spin) channel.}\label{Fig:sus}
\end{figure}

 The most frequently analyzed  quantity in LDA+DMFT studies is the spectral function, i.e., the single particle Green function $A(\mathbf{k},\omega)=-1/\pi\; \text{Im}(G(\mathbf{k},\omega + i0^+)$. This  can be easily computed in LDA+DMFT and directly  compared to (inverse-)photoemission experiments. However, the DMFT analysis can be also extended, by means of an enhanced numerical effort, to the calculation of two-particle quantities.
For understanding the dynamics of the local magnetic moment in the 
iron-pnictides, the quantity of interest is evidently the local spin-spin correlation function $\chi_{\rm loc.}$. In fact, the  significant piece of information enclosed in $\chi_{loc}$  has motivated many experimental groups to measure or estimate accurately $\chi_{loc}$  both in the time- and in the frequency-regime: As we will discuss more extensively in Sec. IV, these experimental estimates represent an important test for the relevance of dynamically screened local magnetic moments in the physics of iron pnictides, as it emerges from our LDA+DMFT calculations.

In the paramagnetic case, when the spin-orientation can be assumed to be isotropic, the local spin correlation function reads
\begin{equation}
\chi_{\rm loc.}(\tau) = \sum_{i,j} \chi_{\rm loc.}^{i,j}(\tau)= g^2 \sum_{i,j} \left\langle {\cal T}_\tau S_z^i(\tau) S_z^j(0) \right\rangle
\label{eq:imp_sus}
\end{equation}
where  $S_z^i(\tau)=1/2(n_\uparrow^i(\tau)-n_\downarrow^i(\tau))$ is the $z-$component of the spin operator of the orbital $i$, expressed in terms of the corresponding density operators $n^{i}_\sigma(\tau)=c^{i\dagger}_\sigma(\tau)c^{i}_\sigma(\tau)$ ($\hbar=1)$. Note that, in DMFT, we will calculate $\chi_{\rm loc.}(\tau)$ from the local susceptibility of the converged DMFT impurity model by means of Hirsch-Fye quantum Monte Carlo\cite{HirschFye}. Hence $\tau$ appearing in Eq. \ref{eq:imp_sus} represents here the imaginary time. For $\tau>0$, we can write explicitly
\begin{multline}
\left\langle S_z^i(\tau) S_z^j(0) \right\rangle \!=\!
\frac{1}{4}[ \sum_{\sigma=\uparrow,\downarrow}\left\langle n_\sigma^i(\tau) n_\sigma^j(0)\right\rangle -\langle n_\sigma^i(\tau) n_{-\sigma}^j(0)\rangle ].
\label{eq:imp_sus2}
\end{multline}
For the non-interacting case or, similarly, for a given quantum Monte-Carlo auxiliary spin configuration \cite{HirschFye} $\{s\}$, this  can be expressed  in terms of the Green function matrix $G^{ij}_{\sigma \{s\}}(\tau,0)=\left\langle c^\dagger_{i,\sigma}(\tau)c_{j,\sigma}(0)\right\rangle_{\{s\}}$ as 
\begin{multline}
\left\langle S_z^i(\tau) S_z^j(0) \right\rangle=
\frac{1}{4}[\sum_{\sigma=\uparrow,\downarrow}
G^{ii}_{\sigma \{s\}}(\tau,\tau)G^{jj}_{\sigma\{s\}}(0,0)
+\\
-G^{ij}_{\sigma\{s\}}(\tau,0) G^{ji}_{\sigma\{s\}}(0,\tau) 
 -G^{ii}_{\sigma\{s\}} (\tau,\tau)G^{jj}_{-\sigma,\{s\}}(0,0)]\text{.}
\label{eq:imp_sus3}
\end{multline}
Let us recall that in the present case we work in a local basis for which the off-diagonal elements of the Green function matrix in the orbital indices vanish within the paramagnetic (PM) phase, at least as far as density-density interaction are considered.

After convergence of the DMFT self-consistent loop, quantity \eqref{eq:imp_sus3} is measured by means of Monte Carlo sampling. 
This procedure represents the most direct way to compute local two-particle correlation functions within DMFT. In fact, being (numerically) exact at the level of the impurity model, it includes automatically all vertex corrections to the bare-bubble spin susceptibility (see Fig.\ 1, for the diagrammatic illustration of $\chi_{\rm loc.}$), without the need of explicitly calculating the local irreducible vertex function $\Gamma$. The explicit calculation of $\Gamma$  would be only needed, in the case one is interested to compute the momentum- and frequency dependent spin susceptibility $\chi({\bf q}, \omega)$ in DMFT\cite{DMFTreview}. While this provides further possibilities for the comparison with experiments, for multiband systems such as pnictides the calculation of  $\chi({\bf q}, \omega)$ in DMFT has been so far  possible\cite{park2011} only by introducing additional approximations (e.g., neglecting all frequency dependencies of the DMFT vertex function $\Gamma$).

As a last step before turning to the discussion of our results, in order to make the comparison with previous or new DMFT calculations easier, let us briefly illustrate the details of the low-energy Hamiltonian we employed for our LDA+DMFT calculations. 

First of all, the band-structure of LaFeAsO has been obtained with the WIEN2k package\cite{wien2k}, adopting the generalized gradient approximation exchange-correlation functional introduced by J. P. Perdew \cite{perdew1996}. As a second step, a Wannier projection on a suitable low-energy basis-set of maximally localized Wannier functions has been performed, using the wien2wannier\cite{W2W} and Wannier90\cite{wannier90} packages. The choice of the most suitable low-energy basis-set for DMFT represents clearly an important point in the procedure, as it is intrinsically related to the value of the Hubbard and Hund's exchange interactions used in the calculations, and also to the well-known problem of double-counting\cite{karolak2010,Aichhorn11} in the LDA+DMFT scheme. 

 In this respect, let us recall  that DMFT calculations performed for low-energy model including only the five $3d$-orbital of Fe run into severe problems connected to the different spread of the resulting Wannier functions. Namely, the difference in the Wannier function spreads renders the standard recipes to neglect all double counting corrections within the $d$-manifold highly questionable. In order to avoid such problems an extension of the basis set for the projection in order taking into account the arsenic and oxygen $p$ degrees of freedom (resulting in the so called $dpp$-models) is a straightforward strategy.  \\

However, when considering the 1111 systems (in our case: LaFeAsO), there is an alternative possibility for the construction of a simple low-energy Hamiltonian, restricted to the $d$-manifold. The additional approximation (first proposed by one of us (R.A.) and H.\ Ikeda in Ref.\ \onlinecite{arita09}) can be understood by realizing that in the Fe-$3d$-manifold, there is specifically one Wannier orbital, namely $3z^2-r^2$, whose localization is much stronger w.r.t. to the other four ($xy$, $x^2-y^2$, $xz$, $yz$)\cite{arita09,hansmann10}. Hence, the local Coulomb interactions for this orbital would be different (namely, larger) than for the other ones, making  no longer justified the neglecting of the double-counting correction terms.
 Fortunately, in the case of LaFeAsO, the band mostly connected to the $3z^2-r^2$ orbital lies below the Fermi level and does not contribute to the Fermi surface. Hence, one can  assume, in fact, that only the remaining four bands are contributing significantly to the response functions under consideration. For this reason, we performed the Wannier projection on the four remaining Fe$3d$ orbitals only, i.e., $xy$, $x^2-y^2$, $xz$, $yz$. This represents the most consistent choice with the common assumption of orbital-independent Hubbard and Hund's exchange interaction parameters (see Eq. \ref{eq:ham} below), and, consequently, with the neglecting of double counting corrections for the $d$-only low-energy Hamiltonian of LaFeAsO.  

After specifying our four-band low-energy model, we consider a density-density type of multi-orbital Hubbard interaction, which in second-quantization reads 
\begin{equation}\label{eq:ham}
H_{\text{int}}=\sum_{i=1}^4 U n_{i\uparrow} n_{i\downarrow}  + \sum_{i \neq j} \sum_{\sigma \sigma'} (V-J\delta_{\sigma \sigma'}) n_{i\sigma} n_{j\sigma'}.
\end{equation}
Note that we set the orbital rotational symmetry relation $V\! = \! U\! - \!2J$ between the (orbital-independent) Hubbard repulsion $U$ for two electrons with opposite spin on the same orbital, the off-diagonal repulsion $V$ between different orbitals $i \! \neq \! j$ and the Hund's coupling $J$. The latter lowers the energy of configurations with parallel spins $\sigma \!=\! \sigma'$. 

The resulting impurity model, associated with DMFT, was solved by means of Hirsch-Fye quantum Monte Carlo (QMC)  \cite{HirschFye}. In agreement with our expectation, the spectral functions and Fermi surfaces obtained from such a four-band model agree much better with the $dpp$-models than the five-band $d$-only model does. Moreover, recently published data on the spin-spin correlation function obtained with $dpp$-models\cite{Aichhorn11} agree well with our previous results and, hence, justify the usage of the four band model also for the two-particle susceptibilities. 

As mentioned also above, the values of the interaction parameters depend on the specific choice of the low-energy model. Specifically, the Hund's coupling $J$, stemming from higher order multipole Coulomb interactions, is more robust against the model choice than the Hubbard $U$ connected to the monopole Coulomb term.
Calculation by means of the constrained random phase approximation (cRPA)\cite{cRPA2004}  for the five-band model give\cite{nakamura2008} an intraorbital Coulomb interaction $U$ of about $2.2 \div 3.3$ eV and a Hund's rule coupling of about $0.3 \div 0.6$ eV. Consequently, we take an average value of $J=0.45$eV while we reduce the value of $U$ to $1.8$eV in order to take into account the screening effects of the eliminated $3z^2-r^2$ states and the slightly larger spread of the Wannier functions (due to the $3z^2-r^2$-tails). 

Let us stress, at the end of this section that the chosen values for the local electronic interaction ($U=1.8$eV, $V=U-2J=0.9$eV, $J=0.45$eV)  yield the experimentally observed renormalization of the single particle spectra and gives also results in good agreement with the $dpp$-model\cite{Aichhorn11}. Note that, as described in our previous Letter, Ref.\ \onlinecite{hansmann10}, performing the LDA+DMFT calculations for an Ising-type Hund's exchange (as in Eq. \ref{eq:ham}) does not lead in the case of the 1111 systems (which are characterized by lower values of the interaction parameters $U$ and $J$) to significant errors, not only for one-particle spectral functions\cite{aichhorn09}), but also for the local spin susceptibility (inset of Fig. 2 of Ref. \onlinecite{hansmann10}).

\section{Temperature dependence}
\label{Sec:Tdep}

\begin{figure}[t!]
\begin{center}
\includegraphics[width=9cm]{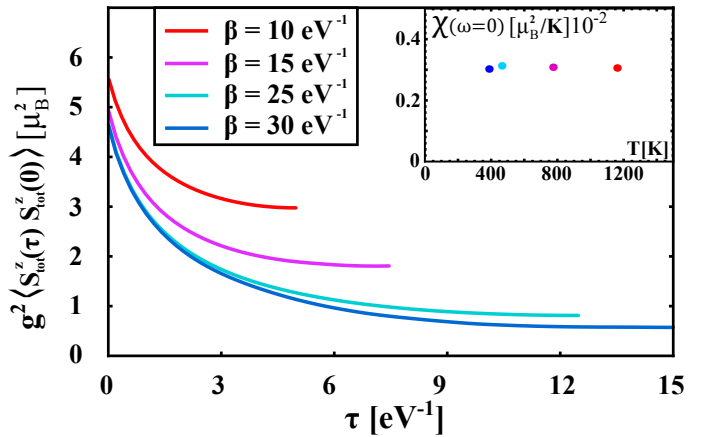}
\end{center}
\vspace{-.5cm}
\caption{(color online) Temperature dependence of the LDA+DMFT  local spin susceptibility as a function of the imaginary time $\tau$. Inset: Temperature dependence of the local static magnetic susceptibility $\chi_{loc}(\omega=0)$, obtained via a $\tau$-integration.}
\label{fig:chi_T}
\end{figure}

In the previous Letter we investigated the dichotomy between the large instantaneous value of the susceptibility $\chi_{loc}(\tau)  = g^2 \langle T_\tau S_z(\tau) S_z(0) \rangle$ at $\tau\!=\!0$ and fixed $T= 460 K$ ($\beta =25$ eV$^{-1}$), which gives us a direct measure of the ``bare'' local moment, and the dynamically screened moment at large $\tau$, which can be related to the moment measured at longer time-scales. 
We discussed the orbital composition of the local spin susceptibility to establish that the inter-orbital Hund's coupling is the key quantity for the observed behavior. 

In order to provide further insight to the picture of magnetic moments driven by local inter-orbital Hund's coupling, which get screened at longer time-scales, we study here also the temperature dependence of $\chi_{loc}(\tau)$. 
In Fig. \ref{fig:chi_T} we show $\chi_{loc}(\tau) $ for different values of $T\!=\!1/\beta$ ranging from $387$K ($\beta= 30$eV$^{-1}$) to 1160K ($\beta=10$eV$^{-1}$).
 From these results, we note a clear saturation of the instantaneous spin-spin correlation function with decreasing temperature to a value more than {\sl twice}as large as the corresponding non-interacting one. 
This means that at $\tau\!=\!0$ a significant local moment is actually formed and it persists at low temperatures. 
The dynamically screened value at $\tau\!=\!\beta/2$, on the other hand,  decreases evidently upon lowering the temperature.  This is consistent with what one expects in a metallic system, i.e. for a Fermi Liquid. Here,  at long enough time scales a complete screening takes places at low-$T$, as a consequence of the strong fluctuations of the local magnetic moment induced by the high electronic mobility (for instance, in the limiting cases of a non-interacting system or of a Fermi Liquid at $T=0$, one finds $\chi(\tau=\beta/2)\!=\!0$). 

Among the various members of the iron-based superconductors, the 1111 family is one of the most itinerant ones, therefore the dynamical screening is quite effective. As a consequence, the difference between the instantaneous and the moment at $\tau\!=\!\beta/2$ is large and the dynamical screening will be effective at shorter time-scales. 

Moreover, the different temperature dependence of  $\chi_{loc}(\tau\!=\!0)$ and $\chi_{loc}(\tau\!=\!\beta/2)$ shown in Fig.~\ref{fig:chi_T} is also reflected in the behavior of the local static susceptibility $\chi_{loc}(\omega\!=\!0) $. In fact, our LDA+DMFT results for $\chi_{loc}(\omega\!=\!0) = \int_0^{\beta} \;  d\tau \; \chi_{loc}(\tau)$, reported in the inset of Fig.\ 2, are very weakly dependent on $T$: We observe essentially a strongly renormalized Pauli behavior. Hence, despite the presence of a large instantaneous magnetic moment, as far as the static moment is concerned, we are very far from the Curie behavior ($ \chi_{loc}(\omega\!=\!0)  \neq \text{const.} \times \frac{1}{T}$), which would require, instead, that $\chi_{loc}(\tau\!=\!0)$ and $\chi_{loc}(\tau\!=\!\beta/2)$ were almost the same (i.e., a screening time-scale going to $\infty$). A Curie-type of behavior is indeed what one obtains for compounds with a higher degree of correlation and localization. For those, because of the poorer metallic screening, a less rapid $\tau$-decay of $\chi_{loc}(\tau)$ is found and the static susceptibility becomes more Curie-like. 

This is likely the reason why the ordered magnetic moment calculated by means of LSDA for some of the most correlated ones (e.g., FeTe) is, strangely enough, in better agreement with experiments than for the less correlated ones. 
In fact, LSDA is a static theory. This means that, even if it cannot reliably describe strongly correlated materials, it may work ``effectively'' better for those compounds in which the poor screening flattens the $\chi_{loc}(\tau)$, i.e., where it becomes less $\tau$-dependent.

Let us note here, for the sake of completeness, that the crossover between Curie-like and (renormalized)  Pauli-like local spin susceptibility has been nicely illustrated by K.~Haule, {\it et al.}\cite{haule09} upon varying the Hund's coupling $J$ for a fixed LDA input. For small values of $J$ (close to that of the 1111 compounds) $\chi_{loc}(\tau)$ is quite constant in $T$ as in our case, which agrees with the eleven band ($dpp$) calculation for LaFeAsO of Ref.\ \onlinecite{Aichhorn11}.  In contrast, for larger values of $J$  the screening of the moments gets less effective, and this results in a pronounced $1/T$ dependence.
The Hund's coupling $J$ is a crucial parameter in driving between these different regimes, in agreement with what has been already put forward by W. Yang {\it et al.} in the context of a x-ray absorption data study \cite{yang09}.
In a more general context, P. Werner,  {\it et al.}\cite{werner2008}, and, afterwords,  L. de' Medici, {\it et al.} \cite{luca} emphasized that the Hund's coupling is a key parameter for tuning correlation effects in multi-band systems, with important differences observed depending on the (integer) filling of the systems\cite{werner2008,luca}.

\section{Frequency dependence}
\label{Sec:omega}

\begin{figure*}[t!]
\begin{center}
\includegraphics[width=18cm]{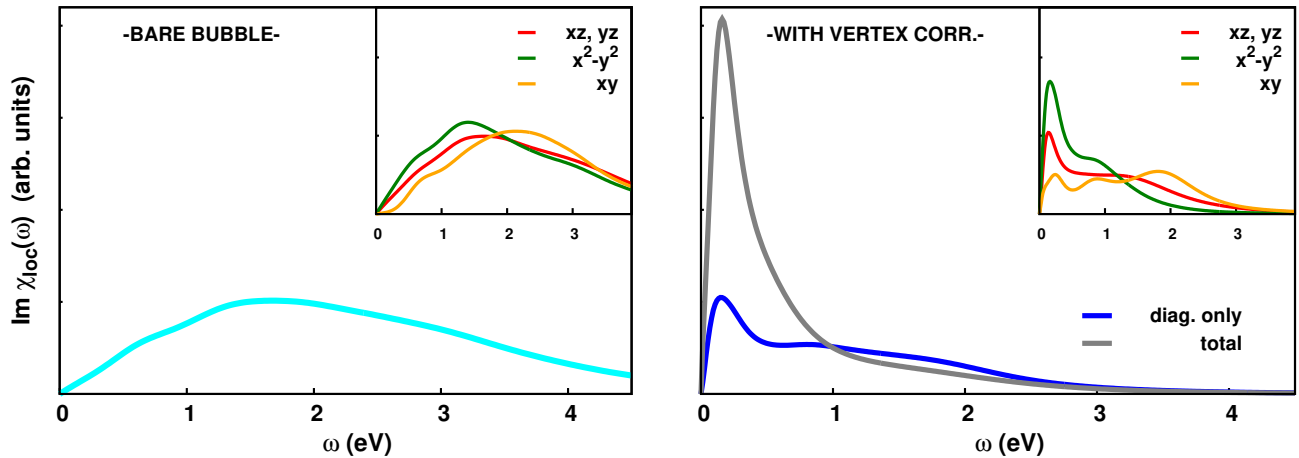}
\end{center}
\vspace{-.5cm}
\caption{Left panel: Frequency dependence of the imaginary part of the local spin susceptibility calculated in LDA+DMFT for LaFaAsO at $\beta=25$eV$^{-1}$ \emph{without} vertex corrections (``bare bubble'' term, see Eq. (\ref{eq:bubble})); inset: orbital resolved contributions to Im$\chi_{loc}(\omega)$ (note that only the diagonal terms are shown, as the non diagonal are identically zero for the bare ``bubble''). Right panel: Frequency dependence of the imaginary part of the local spin susceptibility calculated in LDA+DMFT \emph{with all} vertex corrections included: in blue are shown the sum of all four  diagonal terms, while the sum of all (sixteen) terms is plotted in gray; inset: orbitally resolved plot of the diagonal terms of Im $\chi_{loc}(\omega)$ after the inclusion of vertex corrections.}
\label{fig:chi_w}
\end{figure*}

The calculation of the frequency dependence of the local magnetic spin susceptibility represents an important step forward to relate the ``Hund's metal'' scenario\cite{haule09,yin2} described above with the physics actually observed in iron-based superconductors. In fact, in most of the experiments (among others neutron experiments, see next section) spectral functions (and not time dependent functions) are measured. At the same time, in  LDA+DMFT the analytic continuation of the  local susceptibility $\chi_{loc}(\tau)$ can be obtained by means of the Maximum Entropy Method\cite{MEM} (MEM) directly from the two-particle Green functions of the Anderson impurity model associated with the DMFT, namely from the $\tau$-dependent local magnetic spin susceptibility data, as discussed in the previous section. In comparison with the analytic continuation of the standard one-particle spectral functions, the numerical effort for accumulating high-enough QMC statistic for the MEM is quite larger. Furthermore, the stability of the analytically continued results has been tested by employing different models for the MEM, including also a completely featureless one (constant distribution).        

We start our analysis by considering a rough approximate expression for Im$\chi_{loc}(\omega)$, which is easily obtained by neglecting all vertex corrections ($\Gamma=0$), i.e., by considering only the term with the convolution of two LDA+DMFT retarded Green functions $G_R(\omega)$ (the so-called ``bubble'' term in Fig.\ \ref{Fig:sus}). This approximation, which is well justified for the evaluation of the optical conductivity at the LDA+DMFT level\cite{DMFTreview,tomczak2009,tomczakthesis,PWoptics}, is expected not to work well for the case of spin and charge susceptibility. We discuss first, however, the ``bubble'' results, because often such easy approximation is taken in the theoretical calculations. This preliminary analysis will be also useful in the following for evaluating separately the effects of the vertex corrections in a situation of practical interest, as for the iron-pnictides.

The spectrum of the local magnetic susceptibility Im $\chi_{loc}(\omega)$, computed by considering the ``bubble'' term only, is given by
\begin{equation}
\label{eq:bubble}
\mbox{Im}\, \chi_{loc}(\omega) \! = \! \frac{g^2 \pi}{2} \! \int d\omega' A(\omega') A(\omega'\! +\! \omega)  \left[f(\omega')- f(\omega' \! + \! \omega)\right],
\end{equation}
where $A(\omega)=-\frac{1}{\pi} \mbox{Im} G_{R}(\omega)$ is the local ($\mathbf{k}$-integrated) spectral function obtained by performing the MEM analytic continuation of the local LDA+DMFT Green function, and  $f(x)= 1/(\operatorname{e}^{\beta x}+1)$ is the Fermi-Dirac distribution function.

The LDA+DMFT results for $\mbox{Im} \chi_{loc}(\omega)$ of LaFeAsO at $\beta= 25 $eV$^{-1}$ are shown in Fig.\ \ref{fig:chi_w} (left panel).
The sum of the contributions from all four Fe $d$-Wannier orbitals is shown in the main panel.
In the inset, it can be seen that the orbital dependence is weak. At the level of the ``bubble'' term, the spectrum of the local magnetic susceptibility appears almost featureless, with an unique large maximum, located at $\sim 2 \div 3$eV for all orbitals. This has to be related to the fact that (at least) two important energy scales, i.e. the LDA bandwidths and the local Coulomb interaction in $H_\text{int}$ (Eq. \ref{eq:ham}) are (roughly) of this order of magnitude.

We turn now to evaluate the effects of the inclusion of the vertex corrections in the LDA+DMFT calculations. This corresponds to apply the MEM directly to the two-particle local spin susceptibility $\chi_{loc}(\tau)$ as a function of imaginary time, without resorting to any expression such as Eq.\ (\ref{eq:bubble}) in terms of the single-particle Green function $G_{R}(\omega)$ only. The results of the (bosonic) MEM for $\chi_{loc}(\tau)$ are shown in Fig.\ \ref{fig:chi_w} (right panel), where the sum of all four diagonal contributions (blue line), as well as the total sum of all (sixteen) diagonal and off-diagonal term (solid thick gray line) are reported.  
From a comparison of the total intensity of Im$\chi_{loc}(\omega)$, with the bare ``bubble'' data of Fig.\ \ref{fig:chi_w} (left panel) immediately emerges the important role played by the vertex corrections. They indeed determine the appearance of a well defined peak structure in the spectral functions. Specifically, by looking at the total intensity of Fig. \ref{fig:chi_w} we note a marked low-energy  peak, located at $\omega \sim 0.2$eV, while a second (but much broader) spectral structure is found at larger frequencies.

From the physical point of view, we can look at the development of the low-energy peak as the {\sl fingerprint} of a local moment formation. Such local moment is, however, dynamically screened, as it is witnessed by the relatively visible spectral features at higher frequencies, and -- above all -- by the position of the peak itself, at  $\omega_{peak} \sim  150 \div 200$ meV.  We can better understand this argument, by considering a Fourier analysis of Im$\chi(\omega)$, i.e. its behavior for real times. The peak of Im$\chi(\omega)$ can be directly related to temporal fluctuations of the local spin on the Fe atoms with a characteristic frequency $\sim \omega_{peak}$. These correspond, evidently, to an ``effective'' decay of the local spin moment, when averaging over times larger than the typical fluctuating period, whose size can estimated  ($t_{fluct} \sim \frac{1}{  \omega_{peak}}$) to be of  order of $\sim  10$ femtoseconds (fs). 
  Note that this is larger than the typical time-scales ($t_{el}$) of about $1$ fs characterizing the  electron dynamics in LaFeAsO, i.e. $ t_{el} \sim \frac{1}{Z W} $ (where $W  \sim 4 eV$ is the electronic  bandwidth, and $Z \sim 0.5$ the quasiparticle renormalization factor), which clearly indicates the relevance of the local magnetic moment formation even in the most metallic family (1111)  of the iron-pnictides. At the same time, in contrast to the more correlated case of metals in the proximity of a transition to a Mott insulating phase (where $\omega_{peak} \rightarrow 0$), the local spin fluctuations in LaFeAsO occur  much faster than the  typical time-scales of significant experimental probes, such as  in the case of neutron spectroscopy (see next section for quantitative details of the comparison with experiments).

A second important characteristic sign in Im $\chi(\omega)$ is the importance of the contribution of the off-diagonal terms in the spin-correlation matrix, which is reflected in the difference between the total and the diagonal part of the Im$\chi_{loc}(\omega)$. The significant off-diagonal contribution, which stems entirely from vertex correction effects, can be easily connected with the predominant role of the Hund's exchange interaction in this system.

Finally, as for the diagonal terms, it is also interesting to disentangle our LDA+DMFT data for the local magnetic susceptibility with respect to the four different Fe $d$-orbitals of our model. Contrary to the case of the ``bubble'' approximation (inset of the left panel of Fig. \ref{fig:chi_w}), the spectra (with vertex corrections included) show visible differences among the different orbitals (inset of the right panel of Fig. \ref{fig:chi_w}). In particular, one notes that the first peak structure at low-energy is much more pronounced for the $x^2-y^2$ orbital, i.e., for that whose Wannier function lives most in the plane and whose lobes point in the direction of the ligand As atoms. One also notes the presence of a second peak at energies slightly  smaller than the second maximum in the total (diagonal) spectrum. On the other hand, we note that in the case of the $xy$ orbital (which also lives mainly in the plane, but whose lobes point in the direction of the neighboring Fe atoms), the first peak is considerably weakened. There also appears to be a small shift in frequencies which might however be an artifact stemming from different peak heights, since a combined screening of all orbitals would suggest the first peak to be at the same position. The intensity of the second maximum is also slightly weakened, and the missing spectral weight appears to be shifted towards higher frequency, where a third maximum appears (at $\omega \sim 1.7 \div 1.8$eV). The situation for the degenerate $xz$, $yz$ orbitals is, instead, something in between those of the two planar orbitals, with the exception of the position of the first peak, which appears moving (very slightly) towards lower energies. The emergence of a stronger low-energy peak for the  $x^2-y^2$-orbital appears connected to the value of the orbital occupations, which are respectively $n_{xz}=n_{yz}=1.06$, $n_{x^2-y^2}=0.98$ and $n_{xy}=0.90$. In fact, the $x^2-y^2$ orbital is the ``closest'' to the half-filling ($n=1$) condition, and, hence, one expects more evident effects of strong-correlations (among which the formation of a stronger local magnetic moment).

The evaluation of the vertex correction effects on the LDA+DMFT local magnetic susceptibility allows not only for a more precise description of the physics characterizing the different orbitals and of the most-relevant off-diagonal contributions, but it also represents a necessary step for performing a more quantitative comparison with experiments. In fact, the question of how our theoretical data really compare with spectroscopic experiments in the paramagnetic phase will be addressed extensively in the next section.

\section{Comparison with experiment}
\label{Sec:neutron}

A major outcome of our LDA+DMFT calculation  is the formation of a large instantaneous ($t=0$) local moment, mainly driven by the Hund's exchange interaction $J$ among the four almost degenerate Fe-$3d$ Wannier orbitals at the Fermi level  (Sec. \ref{Sec:Tdep}). However, as the (imaginary) time and frequency dependence (Sec. \ref{Sec:omega}) of the LDA+DMFT local spin susceptibility shows, in the case of LaFeAsO such large local moment is efficiently screened over large time scales  by dynamical fluctuations.  When comparing our LDA+DMFT results with experiments performed in the paramagnetic phase of the Fe-based superconductors, hence, the focus should be on (i) the magnitude of the local magnetic moment at $t=0$ and (ii) its dynamical screening over long time scales.
 
As for (i), one can extract from our LDA+DMFT  $\chi(\tau \! = \! 0)$ value  quantitative estimates for the instantaneous local magnetic moment $m_{loc,t=0}$, and for the ``effective'' magnitude of the total spin $S_{eff}$ of the Fe sites. 
Specifically, after having extrapolated our LDA+DMFT data down to $T= 200$K, and assuming a perfect spin-isotropy for the PM phase, we find $m_{loc,t=0}= g\sqrt{ \langle S_x^2 \rangle +  \langle S_y^2 \rangle +  \langle S_z^2 \rangle} \simeq \sqrt{3\chi_{loc}(\tau=0)} \simeq 3.68 \mu_B$, which would correspond ($m=g \sqrt{S_{eff}(S_{eff}+1)}$) to an ``effective'' spin configuration $S_{eff} \sim 1.4$ for each Fe-atom. Note, that Ising and Heisenberg Hund's exchange yield a similar  $\langle S_z^2 \rangle$ in the  itinerant regime of LaFeAsO \cite{hansmann10}, albeit of course not in more strongly coupled materials.  Such estimates are similar to those ($m_{loc,t=0} \simeq 3.4 \mu_B, S_{eff} \sim 1.3$) obtained in  recent LDA+DMFT calculations\cite{Aichhorn11} for an 11-band $dpp$ model of LaFeAsO. Even higher values for the local magnetic moment were extracted by previous LDA+DMFT calculations ($S_{eff}=1.8$)\cite{craco08}, which  assume a much larger  $U$ and $J$  value than that estimated by cRPA for LaFeAsO.  From the experimental point of view, the detection of a local and instantaneous magnetic moment of the Fe-atoms would evidently require a very fast experimental probe, as, e.g., x-ray absorption spectroscopy (XAS). In fact, XAS measurements of Fe $L_{2,3}$ edge by T. Kroll {\sl et al.}\cite{kroll2010} for LaFeAsO have been fitted by multiplet cluster calculations,  suggesting a high-spin ground state (i.e. $S_{eff}=2$ even larger than the LDA+DMFT predictions).

It is important, however, to test experimentally also the second main aspect of our LDA+DMFT results, i.e., the dynamical screening of such large local magnetic moments over longer time scales. This is possible by inelastic neutron spectroscopy.   
In fact, neutron experiments have been performed for $T > T_N$, e.g., in the case of  CaFe$_2$As$_2$ (at $T=220$K $> T_N= 172$K)  \cite{diallo}
and  (optimally doped) Ba(Fe$_{1-x}$Co)$_2$As$_2$. \cite{lester2010}.
 
 Inelastic neutron spectroscopy are typically re-casted in terms of a {\bf q}- and $\omega$-dependent dynamical structure factor, and this, in turn, can be related to the corresponding ${\bf q}$- and $\omega$- dependent magnetic susceptibility.  Although the crystal field (CF) splitting of iron-pnictides (e.g., CF $\sim 200-300$ meV)  is usually  much lower than the typical values of $1-2$eV of other $3d$ transition metal compounds, in the post-processing of the neutron experimental data for the Fe pnictides the assumption of a {\sl complete} quenching of the orbital moment is usually done. Hence,  the dynamical structure factor is directly related to the imaginary part of the (reduced)\cite{haydenbook} magnetic {\sl spin} susceptibility, i.e., to its spectral function $\mbox{Im} \chi (\bf{q}, \omega)$.

In our case, we are interested in verifying the presence of {\sl local} magnetic moments. Hence, a Fourier transform to real space, i.e., an integral over the momenta {\bf q} is necessary. While neutron scattering data for CaFe$_2$As$_2$ and  Ba(Fe$_{1-x}$Co)$_2$As$_2$  have been taken only for a given set of {\bf q} values, these data have been used for extracting the remaining part of the {\bf q}-dependence of $\mbox{Im} \chi(\mathbf{q}, \omega)$ via a fitting procedure\cite{diallo,lester2010,johnstonrev}, which made it possible to perform the Brillouin zone (BZ) {\bf q}-integration.
On the other hand,  the dynamical screening predicted by our LDA+DMFT calculations is detectable only over a short time scale. This  essentially requires an integral over all frequencies, i.e., a  cut-off $\Omega_C \rightarrow \infty$  in
\begin{eqnarray}
m^2_{loc, t=0} & = &\frac{3}{\pi}\lim_{\Omega_c \rightarrow \infty} \frac{\int_{-\Omega_C}^{\Omega_c} \int_{BZ} \mbox{Im}\, \chi({\bf q},\omega) \, b(\omega) \, d{\bf q} \, d\omega}{\int_{BZ} d {\bf q}} \nonumber \\
&  = & \frac{3}{\pi} \lim_{\Omega_c \rightarrow \infty} \int_{-\Omega_C}^{\Omega_c} \mbox{Im}  \chi_{loc}(\omega) \, b(\omega)  d\omega   .
\label{eq:mom_loc}
\end{eqnarray}
Here $b(x)=1/(\operatorname{e}^{\beta x}-1)$ is the Bose-Einstein distribution function, and the coefficient $3$ comes (as mentioned also above) from the sum over the three spin components ($\langle S^2 \rangle =\langle S_x^2 \rangle  + \langle S_y^2\rangle  + \langle S_z^2 \rangle$ in the PM phase, where we can consider the system to be magnetically isotropic).

The most problematic step is the frequency integral since, e.g., in the case of CaFe$_2$As$_2$ the spin spectral functions have been measured only up $\Omega_C \sim 60 - 80$ meV. One can try to extrapolate the data to higher-frequencies: We consider here explicitly the values obtained\cite{diallo,johnstonrev} for a cut-off $\Omega_C =100$meV, i.e., not too far away from the experimental window.  

\begin{figure}[tb]
\begin{center}
\includegraphics[width=9cm]{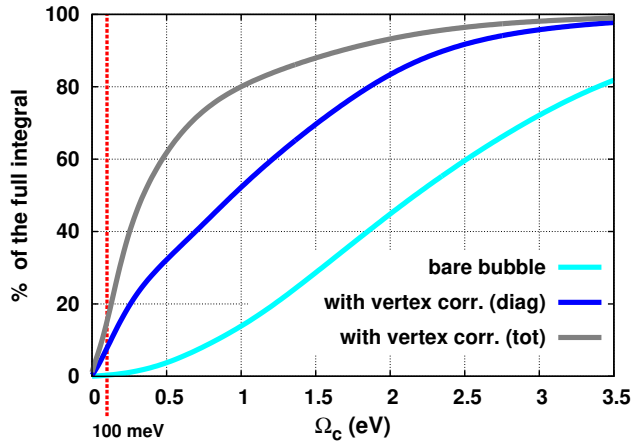}
\end{center}
\vspace{-.5cm}
\caption{(Color online) Cut-off dependence of the frequency integral of Eq. (\ref{eq:mom_loc}) for the {\sl local} (${\bf r}=0$)  and {\sl instantaneous} ($t=0$) magnetic moment calculated from the LDA+DMFT spin-susceptibility \emph{with} (blue line: diagonal terms only; dark gray line: total)  and \emph{without} (light blue line) vertex corrections (i.e., just the bare bubble contribution). The most important contribution to the frequency-integral comes from the energy region  $\omega > 100$ meV, i.e. from higher energies than the typical ones  of the neutron experiments performed for the Fe-based superconductors. If considering a (rounded) typical cut-off of $100$ meV (vertical red dashed line), it is evident that the estimate of Eq. (\ref{eq:mom_loc}) for the {\sl local} and {\sl instantaneous} magnetic moment is much smaller (at least three times) than its ``correct'' asymptotic value.}
\label{fig:cumul_chi}
\end{figure}
With such a cut-off, the corresponding experimental estimate for $m_{loc}(t=0)$ in CaFe$_2$As$_2$ obtained via Eq. \ref{eq:mom_loc} (with an extra factor $\frac{1}{2}$ for obtaining the squared moment per Fe atom\cite{diallo})  was  $m_{loc,t=0} \sim 0.5 \mu_B$ (to be compared with $0.95 \mu_B$ for the ordered moment in the same compound\cite{dialloantro}). Notice that similar value for $m_{loc,t=0}$, and namely  $0.4\mu_B - 0.8 \mu_B$  have been also estimated  for BaFe$_2$As$_2$\cite{matan2009,lester2010} (for the optimally doped compound or in the ordered case, respectively).  

At first glance, hence, one would conclude that the size of the local and instantaneous moment in CaFe$_2$As$_2$ is significantly smaller than that predicted by  the LDA+DMFT calculations (note that for the 122 family, even larger value for $m_{loc,t=0}$ than for LaFeAsO are found\cite{Yin11}). However, one should keep in mind that the LDA+DMFT $\tau=0$ value corresponds  to $\Omega_C=\infty$ in Eq. (\ref{eq:mom_loc}). 

In order to understand to what extent the observed discrepancy might originate from such a cut-off ``mismatch'' between theory ($\Omega_C =+\infty$) and experiment (after a data extrapolation\cite{johnstonrev,diallo} up to $\Omega_C = 100$meV), we can now analyze further the frequency dependence of the local spin susceptibility, which was discussed in the previous section. From the bare bubble contribution to the spin susceptibility (Fig.\ \ref{fig:chi_w}, left panel), it is clear that the major contribution to the integrated spectral weight comes from frequencies larger than $1$eV. Including vertex corrections (Fig.\ \ref{fig:chi_w}, right panel) yields  a richer spectral structure with a prominent peak at about $\omega \sim 0.2$eV. Our theoretical prediction of such a peak is not only consistent with the increase of ${\bf q}$-integrated experimental neutron spectral function for CaFe$_2$As$_2$ and  Ba(Fe$_{1-x}$Co$_x$)$_2$As$_2$ reported in Ref. \onlinecite{lester2010}), but has also been very recently confirmed by novel neutron measurements\cite{premiataditta} in an extended energy range.

 In any case, the spectral weight lying below the experimental cutoff represents just a small fraction of its total value, even the main features of the LDA+DMFT spectral function are located beyond the experimental cut-off. Hence we have plotted in Fig. \ref{fig:cumul_chi} the value of the frequency integral over the LDA+DMFT data up to a cutoff $\Omega_C$:
the experimental cut-off $\Omega_C=100$meV yields only  $10\%$ to  $15\%$ of the instantaneous squared local moment Eq.\ (\ref{eq:mom_loc}).
If assuming that also for  CaFe$_2$As$_2$ only such a fraction of the moment
has been integrated up to $\Omega_C=100$meV, we  get an estimate of $m_{loc,t=0}\sim 1.5 \mu_B$ as the true magnetic moment. This represents of course only
a rough estimate, significantly larger than the neutron spectroscopic values at about $\Omega_C =100$meV but still smaller than the LDA+DMFT/XAS predictions. The remaing discrepancy  might be partly originated  by the lack of SU(2) symmetry of the interaction terms we assumed in Eq. (\ref{eq:ham}), or, more likely, by the effects of spatial correlations neglected by DMFT, which -at present- cannot be easily included via the cluster\cite{CDMFT} or diagrammatic extensions\cite{DGA,DF} of DMFT in complicate systems like the Fe-based superconductors.

\section{Summary and Conclusion}
\label{Sec:Summary}

In this paper, we studied by means of LDA+DMFT the local magnetic properties of iron-based superconductors focusing, in particular, on the 1111 systems (LaFeAsO) for which a small ordered magnetic moment ($0.3-0.6 \mu_B$) is experimentally observed. We have constructed by the Wannier-interpolation technique\cite{wannier90,W2W} an effective low-energy model, which contains the  $d_{xy}$, $d_{x^2-y^2}$, and $d_{yz/zx}$ orbital degrees of freedom, and supplemnted it
with local interactions $U=1.8\,$eV and $J=0.45\,$eV.
 These interaction values are motivated from constrained random phase approximation (cRPA) calculations and reproduce analogous results for the spectral function of the $dpp$ model\cite{miyake10,aichhorn09}:  The system presents an intermediate value of the quasiparticle renormalization factor  $\sim2$ induced by electronic correlations, in agreement to the results of the ARPES experiments.

In order to clarify the physics of local magnetic moments and their dynamical screening as well as to make contact with the experiments, we have analyzed the local magnetic spin susceptibility within the framework of LDA+DMFT and looked into the detail of its temperature and frequency dependence. As for the former, we found the formation of considerably strong {\sl local} ($\mathbf{r}=0$) and {\sl instantaneous} ($t=0$) magnetic moment, which is mainly driven by inter-orbital Hund's coupling $J$, and whose magnitude decays rapidly in  imaginary time. More specifically, the characteristic features of such as ``Hund's metal'' physics are: ($i$) The instantaneously (short-time) local magnetic moment  which corresponds to the square-root of the local spin correlation function $\chi_{loc}(t=0)$ shows a saturation with decreasing temperature.
 ($ii$) The dynamically screened value at long times  ($\tau\!=\!\beta/2$), on the other hand, decreases upon lowering the temperature, as a consequence of the still relatively good metallic properties of the system. 
($iii$) The local  magnetic susceptibility at $\omega=0$, which is the average (integration) over all $\tau$'s,  shows a (strongly renormalized)  Pauli-like behavior in agreement with Ref.\ \onlinecite{Aichhorn11}. For  larger $J$ values, a crossover to a  Curie-like susceptibility is found\cite{haule09}.

A clear hallmark of the physics of the dynamical screening of the large local instantanous magnetic moments has been individuated in the magnetic spectral function, namely in the {\bf q}-integrated spin susceptibilty Im$\chi(\omega)$. In fact, when the vertex correction are properly included, the total spin susceptibility calculated in LDA+DMFT diplays a characteristic, very pronounced peak around $\omega \sim 0.2$eV. 


When comparing our results to experiments, we note that the existence of a large (spin) magnetic moment on the Fe-atoms is supported by x-ray absorption spectroscopy data for LaFeAsO\cite{kroll2010}. The comparison with the neutrons spectroscopy data is more difficult: According to the LDA+DMFT calculations about $90\%$ of the magnetic spectral weight lies above the typical experimental  frequency window  $\sim 60- 80$ meV. This  is compatible with the  observed experimental
increase of  the ${\bf q}$-integrated Im$\chi_{loc}(\omega)$ throughout the measured frequency range,  as well as with the peak structure of the local spin susceptibility in one of the most recent neutron experiment\cite{premiataditta},  and yields a  crude estimate for the instantaneous magnetic moment of at least $\sim 1.5\mu_B$ from the neutron data.  

\vskip 3mm 

We thank  S.\ L.\ Skornyakov, Z.\ Yin,  L.\ De Medici, L.\ Boeri, N.\ Parragh, M.\ Aichhorn, M.\ Capone, V.\ Antropov, D.\ Johnston, G. Kotliar,  and  S. Hayden for fruitful discussions and exchanging of ideas. We acknowledge financial support from the Austrian Science Fund (FWF) through  I610-N16 (AT), JST-PRESTO, Grants in-Aid for Scientific Research from MEXT of Japan (Grant No.23340095)(RA), FWF through F4103-N13 (GS), and Research Unit FOR 1346, Project No.\ I597-N16 of FWF (KH).  Calculations have been performed on the Vienna Scientific Cluster (VSC).



\end{document}